\documentclass[aps,pra,showpacs,floatfix,notitlepage]{revtex4-1}

\usepackage{graphicx,amsmath,amssymb,psfrag,empheq,hyperref,wasysym}
\usepackage{dcolumn}
\usepackage{bm}
\usepackage{epsfig}
\begin{document}

\title{Schr\"odinger Equation for a Non-Relativistic Particle in a Gravitational Field confined by Two Vibrating Mirrors}

\author{Mario Pitschmann}
\email{mario.pitschmann@tuwien.ac.at}
\affiliation{Atominstitut, Technische Universit\"at Wien, Stadionallee 2, A-1020 Wien, Austria}

\author{Hartmut Abele}
\email{hartmut.abele@tuwien.ac.at}
\affiliation{Atominstitut, Technische Universit\"at Wien, Stadionallee 2, A-1020 Wien, Austria}


\begin{abstract}
We derive approximate analytical solutions for a particle in a homogenous gravitational field and confined between two independently vibrating mirrors. This constitutes an extension of the \textit{q}BOUNCE experiment in which ultra-cold neutrons are employed for Gravity Resonance Spectroscopy.
\end{abstract}


\maketitle


\section{Introduction}

We analyze the behavior of a non-relativistic particle, e.g. an ultra-cold neutron, in a homogenous gravitational field and confined between two independently harmonically vibrating mirrors, which constitutes a novel generalization of the \textit{q}BOUNCE experiment. While the \textit{q}BOUNCE setup consists of either one reflecting mirror or two mirrors with constant separation, we analyze here the generalized case with the two mirrors vibrating independently. This adds another layer of complexity in the theoretical description with respect to the former setup. In fact, the \textit{q}BOUNCE experiment constitutes a realization of a quantum mechanical system with time-dependent boundary conditions \cite{Burgan:1979,Morales:1994,Glasser:2009,Fojon:2010,Munier:1981,Martino:2013,Matzkin:2018b}. 

Already in 1949 Fermi \cite{Fermi:1949ee} proposed a mechanism for the production of cosmic rays in which particles move in inhomogeneous magnetic fields. He proposed no specific model but provided some estimates on contemporary data. Later on, Ulam \cite{Ulam:1961} introduced the "Fermi-accelerator" in which a classical ball bounces back and forth between two oscillating walls. His numerical studies showed regular as well as stochastic motion and since then have become a paradigmatic model in chaos studies \cite{Seba:1990,Scheininger:1991,Glasser:2009,Grubelnik:2014}. Soon thereafter the quantum mechanical version emerged in the area of quantum chaos. 
Another line of studies concerning this quantum mechanical system involves non-locality induced by the moving wall on a localized state \cite{Greenberger:1988a,Makowski:1992,Zou:2000,Qian-Kai:2001a,Wang:2008,Mousavi:2012b,Mousavi:2012c,Mousavi:2014a} (for a recent disproof of this conjecture see \cite{Matzkin:2018a}.)
Still another context of this system concerns chirped frequency excitations of quantum states, which were proposed in \cite{Manfredi:2015inn}.
As it turns out, the resolution of the Schr\"odinger equation for an arbitrary time dependent Hamiltonian is a formidable task and beset with conceptual issues.

In this paper we consider the theoretical analysis of such a generalized setup. Our main focus are quantum states of a neutron in Earth's gravitational potential, which have been observed in 2002 \cite{Nesvizhevsky:2002ef} and consecutively in \cite{Nesvizhevsky:2005ss, Nesvizhevsky:2003ww,Abele:2009,Jenke:2009,Jenke:2011,Jenke:2014yel,Cronenberg:2018qxf,Sedmik:2019twj}. A theoretical description can be found in \cite{Voronin:2005xg,Westphal:2007}. The energy eigenstates are discrete and allow to perform precision measurements with a method of resonance spectroscopy as realized by the \textit{q}BOUNCE experiment \cite{Jenke:2011}. In the version of the experiment with Rabi-spectroscopy an energy resolution of  2$\times$10$^{-15}$ peV has been achieved \cite{Cronenberg:2018qxf}.
The experiment was performed in such a way, that ultra-cold neutrons pass three regions, while being reflected on polished glass mirrors. In \cite{Cronenberg:2018qxf}, the resonance spectroscopy transitions between the energy ground state $E_1 = 1.407$ peV and the excited states $E_3 = 3.321$ peV as well as $E_4 = 4.083$ peV have been demonstrated. First, the neutrons pass region I which acts as a state selector for the ground state $|1\rangle$ having energy $E_1$. A polished mirror at the bottom and a rough absorbing scatterer on top at a height of about 20 $\mu$m serve to select the ground state. Neutrons in higher, unwanted states are scattered out of the system. This region has a length of 15 cm. Subsequently, in region II, a horizontal mirror performs harmonic oscillations with a tunable frequency $\omega$, which drives the system into a coherent superposition of ground and excited states. The length of this region is 20 cm. Finally, region III is identical to the first region and hence  acts again as a ground state selector. 

The \textit{q}BOUNCE experiment, has been employed in the search for new physics, i.e. axions and axion-like particles which are candidates of dark matter \cite{Jenke:2014yel}, chameleons and symmetrons which are prominent candidates for screened dark energy \cite{Ivanov:2012cb,Ivanov:2016rfs,Brax:2017hna,Cronenberg:2018qxf} and Lorentz violation \cite{Ivanov:2019}. In this work, we consider the behavior of a neutron, or more generally a non-relativistic particle in only a single region of the \textit{q}BOUNCE experiment but with the second mirror now vibrating independently of the first. At least what concerns the theoretical analysis this leads to a highly  
non-trivial extension of the previous case.

In section \ref{sec:I} the analysis for a particle confined by two independently vibrating mirrors in a gravitational field is detailed, i.e. the corresponding potential operator for the Schr\"odinger equation is derived. Section \ref{sec:II} provides an approximation for the typical case of small mirror vibration amplitudes compared to the mirror separation. Then, section \ref{sec:III} provides the quantum mechanical description of the system under study, while our results are summarized in section \ref{sec:IV}. In Appendices \ref{sec:A} and \ref{sec:B} analytical details relevant for our investigation are presented.

\section{Schr\"odinger Equation for the Neutron in the Gravitational Field}\label{sec:I}

\begin{figure}[h]
\centering
\includegraphics[width=0.4\linewidth]{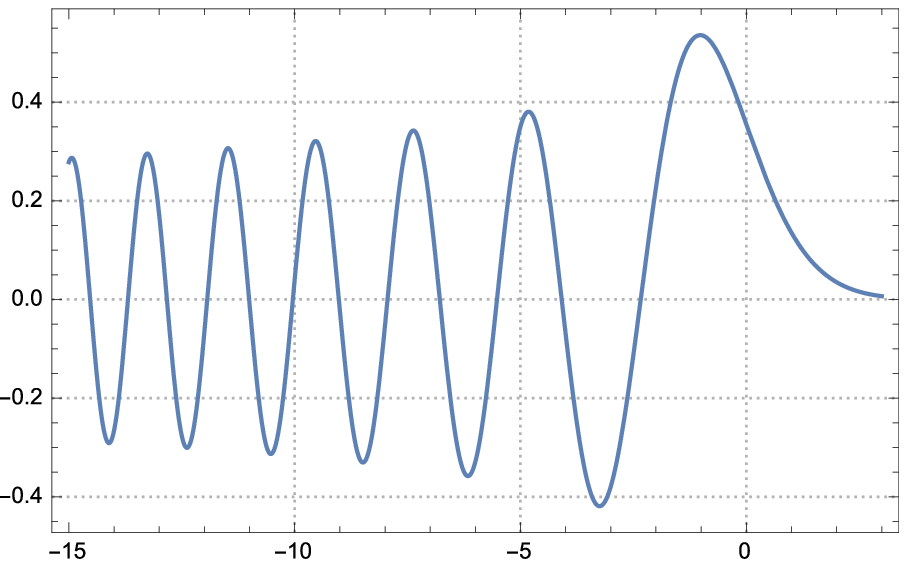} 
\hspace{5mm}
\includegraphics[width=0.4\linewidth]{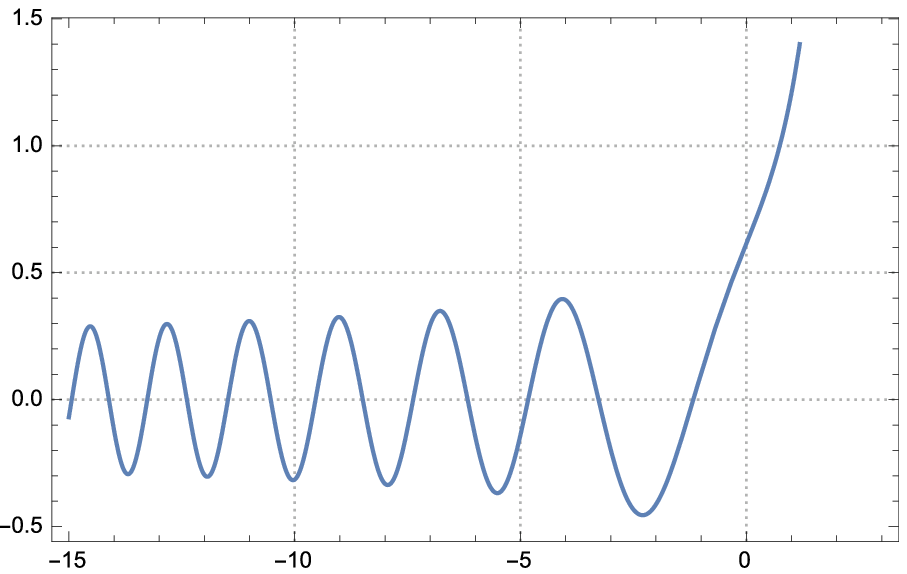} 
\caption{\textit{Left}: Here, the Airy function $\text{Ai}(\sigma)$ is plotted. \textit{Right}: The Airy function $\text{Bi}(\sigma)$ is depicted.}
\label{Fig1}
\end{figure}

\begin{figure}[h]
\centering
\includegraphics[width=0.4\linewidth]{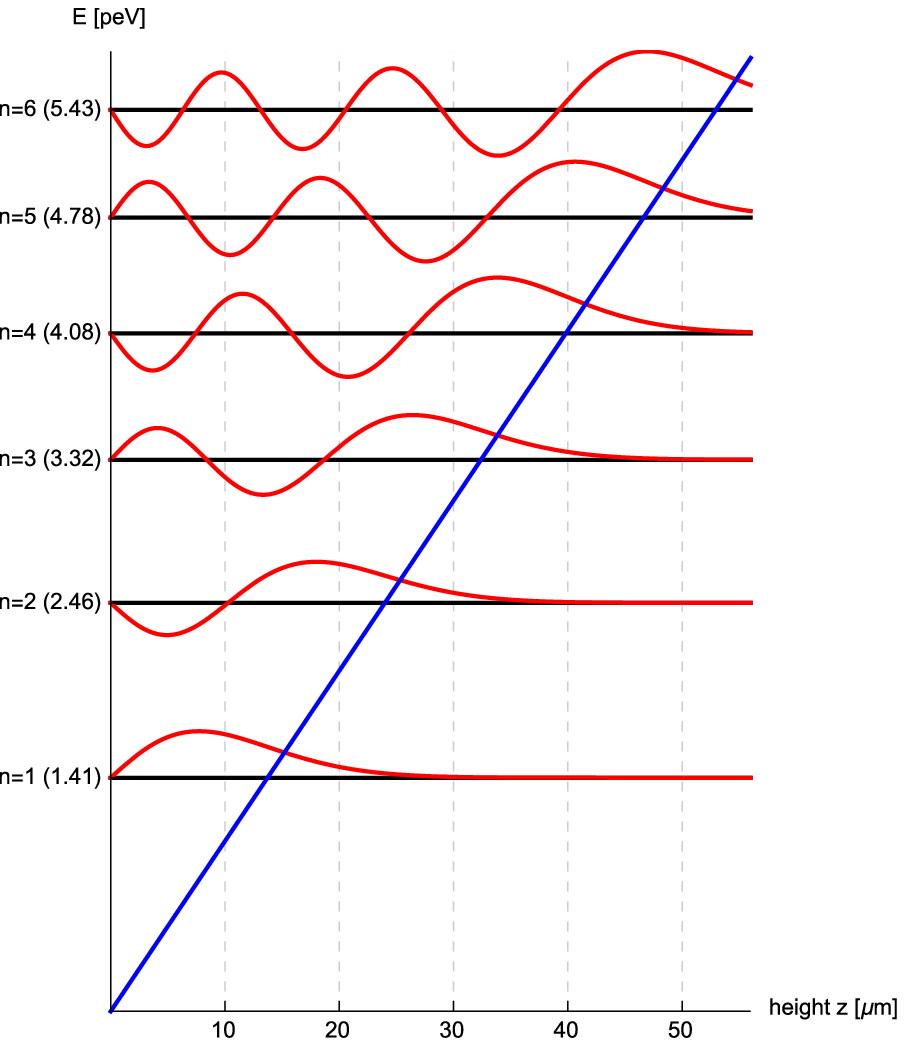} 
\hspace{1cm}
\includegraphics[width=0.4\linewidth]{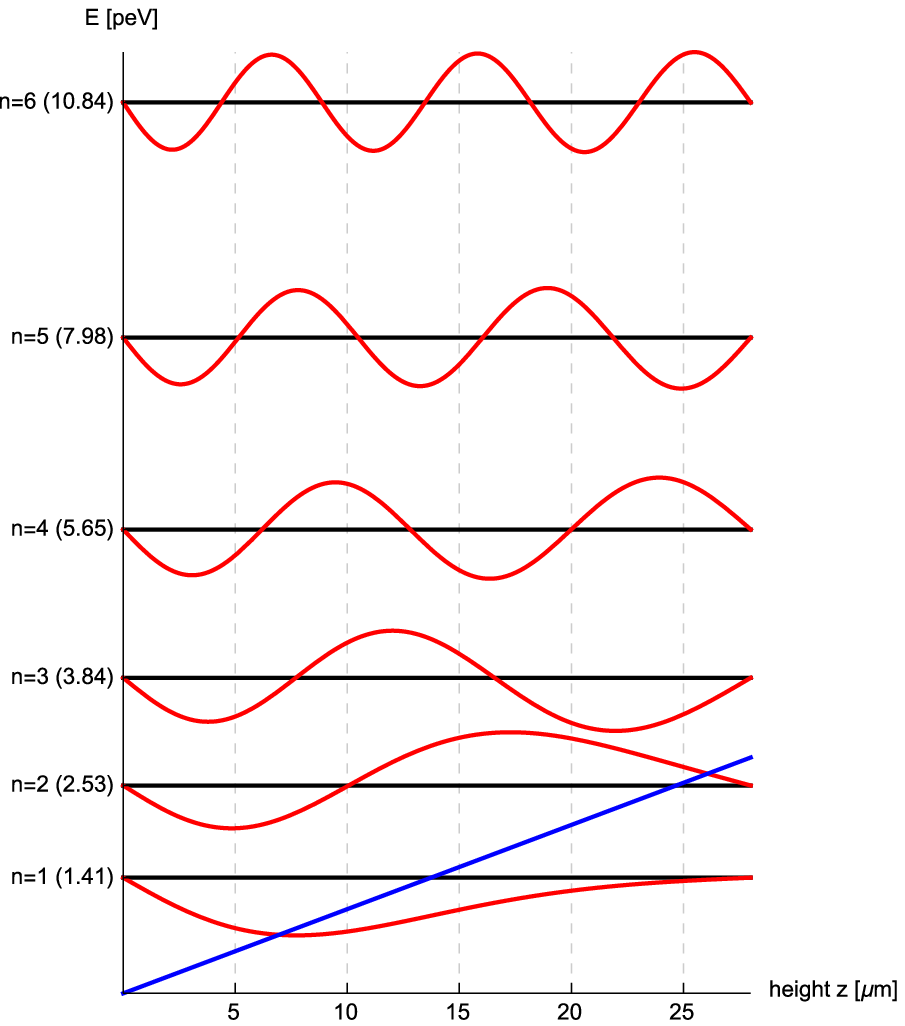} 
\caption{\textit{Left:} The first $6$ wavefunctions are depicted with the gravitational potential $mgz$ as blue line. \textit{Right:} The first $6$ wavefunctions are depicted for $2$ mirrors with separation  $L=28$ $\mu$m with the gravitational potential $mgz$ as blue line.}
\label{Fig2}
\end{figure}

The quantum-mechanical description of an ultra-cold neutron above a mirror in the gravitational potential is given by the Schr\"odinger equation \cite{Westphal:2006dj, Abele:2009dw}. After separation into free transversal and due to gravitation and mirror in vertical $z$ direction bound states, i.e.
\begin{align}
   \Psi_n^{(0)}(\textbf{x},t) = \frac{e^{\frac{i}{\hbar}(p_\perp\cdot x_\perp - E_\perp t)}}{2\pi\hbar v_\perp}\,\psi_n^{(0)}(z)\,e^{-\frac{i}{\hbar}E_n t}\>,
\end{align}
the one-dimensional Schr\"odinger equation for the vertical direction reads 
\begin{align}\label{SEQ}
   -\frac{\hbar^2}{2m}\frac{\partial^2\psi_n^{(0)}(z)}{\partial z^2} + mgz\,\psi_n^{(0)}(z) = E_n\psi_n^{(0)}(z)\>,
\end{align}
where $m$ is the mass of the particle (neutron) and $g$ the local gravitational acceleration. The index $^{(0)}$ denotes wavefunctions unperturbed by any mirror movement. The characteristic length scale and energy scale are given by 
\begin{align}
  z_0 = \sqrt[3]{\frac{\hbar^2}{2m^2g}} = 5.87\,\mu\text{m}\>, \qquad E_0 = \sqrt[3]{\frac{\hbar^2 mg^2}{2}} = 0.60165\text{ peV}\>.
\end{align}
With the substitution of
\begin{align}
  \sigma = \sqrt[3]{\frac{2m^2g}{\hbar^2}}\left(z - \frac{E_n}{mg}\right) \equiv \frac{z - z_n}{z_0}\>,
\end{align}
Eq.~(\ref{SEQ}) reduces to Airy's equation
\begin{align}
   \frac{d^2\psi_n^{(0)}(\sigma)}{d\sigma^2} - \sigma\,\psi_n^{(0)}(\sigma) = 0\>.
\end{align}
The two linearly independent solutions are the Airy functions $\text{Ai}(\sigma)$ and $\text{Bi}(\sigma)$ with the integral representations (see e.g. \cite{Vallee:2004})
\begin{align}
   \text{Ai}(\sigma) &= \frac{1}{\pi}\int_0^\infty \cos\bigg(\frac{t^3}{3} + \sigma t\bigg) dt\>, \nonumber\\
   \text{Bi}(\sigma) &= \frac{1}{\pi}\int_0^\infty \bigg[\exp\bigg(-\frac{t^3}{3} + \sigma t\bigg) + \sin\bigg(\frac{t^3}{3} + \sigma t\bigg)\bigg] dt\>.
\end{align}
They are plotted in Fig.~\ref{Fig1}.
The asymptotic forms are 
\begin{align}
\lim_{\sigma\to\infty} \text{Ai}(\sigma) &\simeq \frac{e^{-\frac{2}{3}\sigma^{3/2}}}{2\sqrt\pi\,\sigma^{1/4}}\qquad, \qquad \lim_{\sigma\to-\infty} \text{Ai}(\sigma) \simeq -\frac{\sin(\frac{2}{3}|\sigma|^{3/2} + \frac{\pi}{4})}{\sqrt\pi\,|\sigma|^{1/4}}\>, \nonumber\\
\lim_{\sigma\to\infty} \text{Bi}(\sigma) &\simeq \frac{e^{\frac{2}{3}\sigma^{3/2}}}{\sqrt\pi\,\sigma^{1/4}}\qquad \hspace{1.5mm}, \qquad \lim_{\sigma\to-\infty} \text{Bi}(\sigma) \simeq -\frac{\cos(\frac{2}{3}|\sigma|^{3/2} + \frac{\pi}{4})}{\sqrt\pi\,|\sigma|^{1/4}}\>.
\end{align}
In the case of $1$-mirror we have the normalized solution (see Eq.~(\ref{AppNWF}))
\begin{align}\label{1m}
   \psi_n^{(0)}(z,t) = \frac{1}{\displaystyle\sqrt{z_0}}\frac{\displaystyle\text{Ai}\Big(\frac{z - z_n}{z_0}\Big)}{\text{Ai}'\Big(-\displaystyle\frac{z_n}{z_0}\Big)}\,e^{-\frac{i}{\hbar} E_n t}\>.
\end{align}
The discrete energy spectrum is obtained from the condition that the wavefunctions vanish at the mirror surface, i.e.
\begin{align}
   \psi_n^{(0)}(0) \propto \text{Ai}\bigg(-\sqrt[3]{\frac{2}{mg^2\hbar^2}}\,E_n\bigg) = 0\>.
\end{align}
The first six eigenvalues obtained are listed in table~\ref{tab:1}
\begin{table}[h]  
\centering
\addtolength{\tabcolsep}{2pt}
\renewcommand{\arraystretch}{1.4}
\begin{tabular}{|c||c|}
  \hline
  $n$ & $E_n$ [peV]   \\
  \hline\hline
   1 & 1.40672 \\
   \hline
   2 & 2.45951 \\
   \hline
   3 & 3.32144 \\
   \hline
   4 & 4.08321 \\
   \hline
   5 & 4.77958 \\
   \hline
   6 & 5.42846 \\
  \hline
\end{tabular}
\caption{The first six energy eigenvalues corresponding to the 1-mirror wavefunctions Eq.~(\ref{1m}) are listed above.}
\label{tab:1}
\end{table}
with the corresponding eigenfunctions depicted in Fig.~\ref{Fig2} \textit{left}.
In the case of $2$-mirrors with fixed separation $L$ we find the normalized solution having $m - 1$ nodes (see Eq.~(\ref{AppNWF}))
\begin{align}\label{WF2MU}
&\psi_m^{(0)}(z,t) = \nonumber\\
&\frac{1}{\sqrt{z_0}}\frac{\displaystyle\Big\{\text{Bi}\Big(\frac{- \bar z_m}{z_0}\Big)\,\text{Ai}\Big(\frac{z - \bar z_m}{z_0}\Big) - \text{Ai}\Big(\frac{- \bar z_m}{z_0}\Big)\,\text{Bi}\Big(\frac{z - \bar z_m}{z_0}\Big)\Big\}\,e^{-\frac{i}{\hbar} \bar E_m t}}{\displaystyle\sqrt{\Big\{\text{Bi}\Big(\frac{- \bar z_m}{z_0}\Big)\,\text{Ai}'\Big(\frac{- \bar z_m}{z_0}\Big) - \text{Ai}\Big(\frac{- \bar z_m}{z_0}\Big)\,\text{Bi}'\Big(\frac{- \bar z_m}{z_0}\Big)\Big\}^2 - \Big\{\text{Bi}\Big(\frac{- \bar z_m}{z_0}\Big)\,\text{Ai}'\Big(\frac{L - \bar z_m}{z_0}\Big) - \text{Ai}\Big(\frac{- \bar z_m}{z_0}\Big)\,\text{Bi}'\Big(\frac{L - \bar z_m}{z_0}\Big)\Big\}^2}}\>,
\end{align}
where the prime denotes derivatives with respect to the argument, i.e. $z_0\,d/dz$. The bars on $z_m$ and $E_m$ are just a reminder that the corresponding numerical values differ in the $1$-mirror and $2$-mirror case.
The energy spectrum is obtained by the conditions that the wavefunctions vanish at the lower as well as upper mirror surface, i.e.
\begin{align}
   \psi_n^{(0)}(0) = \psi_n^{(0)}(L) = 0\>.
\end{align}
For a mirror displacement of e.g. $L=28$ $\mu$m, which is typical of the \textit{q}BOUNCE experiment, the first six eigenvalues are listed in table~\ref{tab:2} and 
\begin{table}[h]  
\centering
\addtolength{\tabcolsep}{2pt}
\renewcommand{\arraystretch}{1.4}
\begin{tabular}{|c||c|}
  \hline
  $n$ & $E_n$ [peV]   \\
  \hline\hline
   1 & 1.40789 \\
   \hline
   2 & 2.52995 \\
   \hline
   3 & 3.84063 \\
   \hline
   4 & 5.64543 \\
   \hline
   5 & 7.98135 \\
   \hline
   6 & 10.8436 \\
  \hline
\end{tabular}
\caption{The first six energy eigenvalues corresponding to the 2-mirror wavefunctions Eq.~(\ref{WF2MU}) with a mirror separation of $L=28$ $\mu$m are listed above.}
\label{tab:2}
\end{table}
the corresponding wavefunctions are depicted in Fig.~\ref{Fig2} \textit{right}.

\subsection{One Vibrating Mirror}

Here, we consider the effect of a single vibrating lower mirror on the particle. Such a configuration has been experimentally realized in region II of the Rabi-spectroscopy version of the \textit{q}BOUNCE experiment. It will be described via a transition to the non-inertial frame of the moving mirror.

Throughout this section, we denote the coordinates of the \textit{inertial frame} as $(z_0, t_0)$ and the wavefunction as $\psi(z_0,t_0)$.  
In the \textit{non-inertial frame} in which the lower mirror is at rest we have coordinates $(\tilde z,\tilde t)$ for which $z_0 = \tilde z + a\sin\omega\tilde t$. Here, $a$ is the amplitude of vibration of the lower mirror and $\omega$ the corresponding vibration frequency. 

The transition to this \textit{non-inertial frame} can be described by the transformation $(z_0, t_0) \longrightarrow (\tilde z,\tilde t)$ and $\psi(z_0,t_0) = \psi(z_0(\tilde z,\tilde t),t_0(\tilde z,\tilde t)) \equiv \tilde\psi(\tilde z,\tilde t)$ with 
\begin{align}
   \tilde t = t_0\>, \qquad
   \tilde z = z_0 - a\sin\omega t_0\>,
\end{align}
and for the derivative operators
\begin{align}
   \frac{\partial}{\partial t_0} &= \frac{\partial}{\partial\tilde t} - a\omega\cos\omega\tilde t\,\frac{\partial}{\partial\tilde z}\>, \nonumber\\
   \frac{\partial}{\partial z_0} &= \frac{\partial}{\partial\tilde z}\>.
\end{align}
Using this in the Schr\"odinger equation
\begin{align}
   -\frac{\hbar^2}{2m}\frac{\partial^2\psi(z_0,t_0)}{\partial z_0^2} + mgz_0\,\psi(z_0,t_0) = i\hbar\,\frac{\partial\psi(z_0,t_0)}{\partial t_0}\>,
\end{align}
we obtain the Schr\"odinger equation in the \textit{non-inertial frame}
\begin{align}\label{Schr1M}
\Big(-\frac{\hbar^2}{2m}\frac{\partial^2}{\partial\tilde z^2} + mg\tilde z + mga\sin\omega\tilde t + i\hbar a\omega\cos\omega\tilde t\,\frac{\partial}{\partial\tilde z}\Big)\,\tilde\psi(\tilde z,\tilde t) = i\hbar\,\frac{\partial\tilde\psi(\tilde z,\tilde t)}{\partial\tilde t}\>.
\end{align}
Hence, the transition from the \textit{inertial frame} to the \textit{non-inertial frame} can be described by the introduction of the additional potential operator
\begin{align}\label{EP1M}
  \hat V =  mga\sin\omega\tilde t + i\hbar a\omega\cos\omega\tilde t\,\frac{\partial}{\partial\tilde z}\>.
\end{align}

\subsection{Addition of a Second Independently Vibrating Mirror}

Next, we consider the addition of a second upper mirror vibrating independently. Such a configuration has not yet been realized in an experimental setup. For the analytical description of the time-dependent mirror separation we follow the procedure in \cite{Munier:1981}, which allows to translate the time-dependence of the mirror separation into a potential operator. Effectively, the problem is thereby reduced to a particle in a homogenous gravitational field confined by two static mirrors. 

The position of the upper mirror in the \textit{inertial frame} shall be given by $z_0^M = L + A\sin(\Omega t_0 + \varphi)$ and in the \textit{non-inertial frame} by $\tilde z^M = L + A\sin(\Omega\tilde t + \varphi) - a\sin\omega\tilde t =: \tilde\chi(\tilde t/T)L$ with the dimensionless factor
\begin{align}
   \tilde\chi(\tilde t/T) := 1 + \frac{A}{L}\,\sin(\Omega\tilde t + \varphi) - \frac{a}{L}\,\sin\omega\tilde t\>.
\end{align}
Here, $a$ denotes again the amplitude of vibration of the lower mirror, while $A$ is the amplitude of vibration of the upper mirror, $\omega$ is the vibration frequency of the lower mirror and $\Omega$ the corresponding vibration frequency of the upper mirror, $L$ the distance between lower and upper mirror at their respective average position, $\varphi$ the phase advance of the upper mirror with respect to the lower mirror at $\tilde t=0$ and $T$ a unit of time. Following \cite{Munier:1981}, we introduce the transformed coordinates
\begin{align}
   z := \frac{\tilde z}{\tilde\chi(\tilde t/T)}\>,  \qquad
   t := \int_0^{\tilde t}\frac{d\tilde t'}{\tilde\chi^2(\tilde t'/T)}\>.
\end{align}
Next, we define
\begin{align}
   \chi(t/T) &:= \tilde\chi(\tilde t/T)\>,  \nonumber\\
   \psi(z,t) &:= \tilde\psi(\tilde z,\tilde t)\>.
\end{align}
The inverse transformations are clearly given by
\begin{align}
   \tilde z = \chi(t/T)\,z\>,  \qquad
   \tilde t = \int_0^t dt'\,\chi^2(t'/T)\>.
\end{align}
Furthermore, we have the relations between the differential operators 
\begin{align}
   \frac{\partial}{\partial\tilde z} &= \frac{1}{\chi(t/T)}\frac{\partial}{\partial z}\>,  \nonumber\\
   \frac{\partial}{\partial\tilde t} &= - \frac{\chi'(t/T)}{\chi^3(t/T)}\frac{z}{T}\frac{\partial}{\partial z} + \frac{1}{\chi^2(t/T)}\frac{\partial}{\partial t}\>,
\end{align}
where a prime $'$ denotes a derivative with respect to the argument. The result of this procedure is that the boundary conditions due to the two vibrating mirrors become those of two static walls with separation $L$ in the new "unbarred" coordinates. Hence, the time dependence of the boundary conditions has been transformed into an additional potential operator in the Schr\"odinger equation.

In transformed coordinates the Schr\"odinger equation Eq.~(\ref{Schr1M}) takes the form
\begin{align}
   \bigg\{-\frac{\hbar^2}{2m}\frac{\partial^2}{\partial z^2} &+ mg\chi^3(t/T)z + mga\chi^2(t/T)\sin\left(\omega\int_0^{t} dt'\,\chi^2(t'/T)\right) \nonumber\\
   &+ i\hbar a\omega\chi(t/T)\cos\left(\omega\int_0^{t} dt'\,\chi^2(t'/T)\right)\frac{\partial}{\partial z} + i\hbar\,\frac{\chi'(t/T)}{\chi(t/T)}\frac{z}{T}\frac{\partial}{\partial z}\bigg\}\,\psi(z,t) = i\hbar\,\frac{\partial\psi(z,t)}{\partial t}\>.
\end{align}
The corresponding Hamiltonian 
\begin{align}
   \mathcal H = -\frac{\hbar^2}{2m}\frac{\partial^2}{\partial z^2} &+ mg\chi^3(t/T)z + mga\chi^2(t/T)\sin\left(\omega\int_0^{t} dt'\,\chi^2(t'/T)\right) \nonumber\\
   &+ i\hbar a\omega\chi(t/T)\cos\left(\omega\int_0^{t} dt'\,\chi^2(t'/T)\right)\frac{\partial}{\partial z} + i\hbar\,\frac{\chi'(t/T)}{\chi(t/T)}\frac{z}{T}\frac{\partial}{\partial z}\>,
\end{align}
is indeed Hermitian, i.e.
\begin{align}
   \langle\mathcal H\rangle = \int_0^L dz\sqrt{-g_{(3)}}\,\psi^*(z)
  \mathcal H\psi(z) = \langle\mathcal H^\dagger\rangle\>,
\end{align}
with the 3-dimensional metric
\begin{align}
  g_{ij}  = \begin{pmatrix}
    1 & 0 & 0 \\
    0 & 1 & 0 \\
    0 & 0 & \chi^2
\end{pmatrix}\>,
\end{align}
and $\sqrt{g^{(3)}}=\sqrt{\det g_{ij}} = \chi$. 

It is more convenient to absorb the Jacobian in the wavefunction, hence we introduce the new wavefunction
\begin{align}
   \Psi= \big(g^{(3)}\big)^{1/4}\psi\>.
\end{align}
Then, the Schr\"odinger equation 
\begin{align}
   \mathcal H\big(g^{(3)}\big)^{-1/4}\Psi&= i\hbar\,\frac{\partial}{\partial t}\left\{\big(g^{(3)}\big)^{-1/4}\Psi\right\} \nonumber\\
   &=i\hbar\left\{\frac{\partial\big(g^{(3)}\big)^{-1/4}}{\partial t}\,\Psi + \big(g^{(3)}\big)^{-1/4}\,\frac{\partial\Psi}{\partial t}\right\},
\end{align}
takes the canonical form
\begin{align}
   \mathfrak{H}\Psi= i\hbar\,\frac{\partial\Psi}{\partial t}\>,
\end{align}
with the new Hamiltonian
\begin{align}
  \mathfrak{H} &= \big(g^{(3)}\big)^{1/4}\,\mathcal H\big(g^{(3)}\big)^{-1/4} - i\hbar\,\big(g^{(3)}\big)^{1/4}\,\frac{\partial}{\partial t}\,\big(g^{(3)}\big)^{-1/4} \nonumber\\
  &= \mathcal H + i\hbar\,\frac{\chi'(t/T)}{\chi(t/T)}\frac{1}{2T}\>.
\end{align}
Explicitly, the Hamiltonian is given by 
\begin{align}
   \mathfrak{H} &= -\frac{\hbar^2}{2m}\frac{\partial^2}{\partial z^2} + mg\chi^3(t/T)z + mga\chi^2(t/T)\sin\left(\omega\int_0^{t} dt'\,\chi^2(t'/T)\right) \nonumber\\
   &\qquad+ i\hbar a\omega\chi(t/T)\cos\left(\omega\int_0^{t} dt'\,\chi^2(t'/T)\right)\frac{\partial}{\partial z} + i\hbar\,\frac{\chi'(t/T)}{\chi(t/T)}\frac{1}{T}\left(z\,\frac{\partial}{\partial z} + \frac{1}{2}\right)\>.
\end{align}
By subtraction of the Hamiltonian for a particle in a homogenous gravitational field without vibrating mirrors, i.e. the Hamiltonian of Eq.~(\ref{SEQ}), we finally obtain the potential operator, which accounts for the independent vibrations of both mirrors 
\begin{align}
  \hat V &= mg\left(\chi^3(t/T) - 1\right)z + mga\chi^2(t/T)\sin\left(\omega\int_0^{t} dt'\,\chi^2(t'/T)\right) \nonumber\\
   &\quad+ i\hbar a\omega\chi(t/T)\cos\left(\omega\int_0^{t} dt'\,\chi^2(t'/T)\right)\frac{\partial}{\partial z} + i\hbar\,\frac{\chi'(t/T)}{\chi(t/T)}\frac{1}{T}\left(z\,\frac{\partial}{\partial z} + \frac{1}{2}\right)\>.
\end{align}
It is convenient to separate this potential into terms, which can induce transitions between quantum mechanical eigenstates $\hat V_T$ and a potential $\hat V_R$ containing the remaining terms, i.e.
\begin{align}
  \hat V = \hat V_T + \hat V_R\>,  
\end{align}
where
\begin{align}\label{EP2M}
  \hat V_T &= mg\left(\chi^3(t/T) - 1\right)z + i\hbar a\omega\chi(t/T)\cos\left(\omega\int_0^{t} dt'\,\chi^2(t'/T)\right)\frac{\partial}{\partial z} + i\hbar\,\frac{\chi'(t/T)}{\chi(t/T)}\frac{1}{T}\left(z\,\frac{\partial}{\partial z} + \frac{1}{2}\right)\>,  \nonumber\\
  \hat V_R &= mga\chi^2(t/T)\sin\left(\omega\int_0^{t} dt'\,\chi^2(t'/T)\right)\>.
\end{align}
In its exact form the Schr\"odinger equation with the potential operators $\hat V_T$ and $\hat V_R$ is not yet amenable for practical evaluations. Therefore, we provide a perturbative approximation in the following section.

\section{A Perturbative Approximation}\label{sec:II}

In this section, we perform a perturbative approximation by considering $A/L\ll 1$ and $a/L\ll 1$, i.e. we consider the amplitudes of the mirror vibrations as small compared to the separation between the mirrors. Consequently, we approximate
\begin{align}
   t &:= \int_0^{\tilde t}\frac{d\tilde t'}{\tilde\chi^2(\tilde t'/T)}  \nonumber\\
   &= \int_0^{\tilde t}\frac{d\tilde t'}{\displaystyle\left(1 + \frac{A}{L}\sin(\Omega\tilde t' + \varphi) - \frac{a}{L}\sin\omega\tilde t'\right)^2}\>.
\end{align}
A Taylor expansion and term to term integration gives 
\begin{align}
   t \simeq \tilde t + \frac{2A}{\Omega L}\cos(\Omega\tilde t + \varphi) - \frac{2A}{\Omega L}\cos\varphi - \frac{2a}{\Omega L}\cos\omega\tilde t + \frac{2a}{\Omega L}\>,
\end{align}
with the corresponding inverse relation to leading order  
\begin{align}
   \tilde t \simeq t - \frac{2A}{\Omega L}\cos(\Omega t + \varphi) + \frac{2A}{\Omega L}\cos\varphi + \frac{2a}{\Omega L}\cos\omega t - \frac{2a}{\Omega L}\>.
\end{align}
For the approximation of $\chi(t/T)$ we find
\begin{align}
   \chi(t/T) = \sqrt{\frac{d\tilde t}{dt}} \simeq 1 + \frac{A}{L}\sin(\Omega t + \varphi) - \frac{a}{L}\sin\omega t\>.
\end{align}
Using these approximations the potential operator describing the vibrations of both mirrors Eq.~(\ref{EP2M}) takes the form
\begin{align}\label{POP1}
  \hat V_T &= mg\left\{\frac{3A}{L}\sin(\Omega t + \varphi) - \frac{3a}{L}\sin\omega t\right\}z + i\hbar a\omega\,\bigg\{\cos\omega t + \frac{A}{L}\cos\omega t\sin(\Omega t + \varphi) - \frac{a}{L}\cos\omega t\sin\omega t \nonumber\\
  &\quad + \frac{2A}{L}\frac{\omega}{\Omega}\sin\omega t\,\big(\cos(\Omega t + \varphi) - \cos\varphi\big) - \frac{2a}{L}\sin\omega t\,\big(\cos\omega t - 1\big)\bigg\}\,\frac{\partial}{\partial z} \nonumber\\
  &\quad + i\hbar\,\left\{\frac{A}{L}\,\Omega\cos(\Omega t + \varphi) - \frac{a}{L}\,\omega\cos\omega t\right\}\left(z\,\frac{\partial}{\partial z} + \frac{1}{2}\right)\>,   
\end{align}
for the part of the potential operator, which can induce transitions between quantum mechanical eigenstates and 
\begin{align}\label{POP2}
  \hat V_R &= mga\,\bigg\{\sin\omega t + \frac{2A}{L}\sin\omega t\sin(\Omega t + \varphi) - \frac{2a}{L}\sin^2\!\omega t \nonumber\\
  &\quad - \frac{2A}{L}\frac{\omega}{\Omega}\cos\omega t\,\big(\cos(\Omega t + \varphi) - \cos\varphi\big) + \frac{2a}{L}\cos\omega t\,\big(\cos\omega t - 1\big)\bigg\}\>,
\end{align}
for the remaining potential operator. 

The special case of the upper mirror being in relative alignment with the lower mirror is described by $A = a$, $\Omega = \omega$ and $\varphi = 0$ in which case both operators reduce 
to the operators of a setup with only a single lower mirror Eq.~(\ref{EP1M}) as expected.

\section{Quantum Mechanical Description of the System}\label{sec:III}

The potential operators describing the independent vibrations of both mirrors in the perturbative approximation, as given in section \ref{sec:II}, are amenable to a standard quantum-mechanical perturbative treatment (see e.g. \cite{Landau:1986}). 
For the wavefunction we write
\begin{align}
  \Psi(z,t) = \sum_k a_k(t)\,\psi_k^{(0)}(z,t)\>,
\end{align}
where $\psi_k^{(0)}(z,t)$ are the eigenfunctions of the unperturbed problem, i.e. for a particle in a homogenous gravitational field and confined by two static mirrors as given in Eq.~(\ref{WF2MU}).
Using the orthogonality of the unperturbed wavefunctions, we find
\begin{align}\label{SDE}
  i\hbar\,\frac{d a_m(t)}{dt} = \sum_k a_k(t)\,\int_0^L dz\,\psi_m^{(0)*}(z,t)\hat V\psi_k^{(0)}(z,t)\>.
\end{align}
This coupled system of differential equations is most conveniently treated numerically, while the individual matrix elements 
\begin{align}\label{ME}
  \int_0^L dz\,\psi_m^{(0)*}(z)\hat V\psi_k^{(0)}(z)\>,
\end{align}
can be evaluated analytically exact. We refrain from reproducing the complicated expressions here, since
this can be straightforwardly done by employing the explicit expressions for the potential operator, i.e. Eqs.~(\ref{POP1}) and (\ref{POP2}), in Eq.~(\ref{ME}). The spatial integration of the corresponding operators 
$z$, $\partial/\partial z$ and $z\,\partial/\partial z$ between two eigenfunctions are derived in detail in the Appendices with the relevant results given in Eqs.~(\ref{App:z1}), (\ref{App:z2}), (\ref{App:dz1}), (\ref{App:dz2}), (\ref{App:zdz1}) and (\ref{App:zdz2}).
In order to relate the more general results from these Appendices to our specific case, we need to employ the specifications 
\begin{align}
  F'(\sigma_A) &= \text{Bi}\Big(\frac{- \bar z_m}{z_0}\Big)\,\text{Ai}'\Big(\frac{ - \bar z_m}{z_0}\Big) - \text{Ai}\Big(\frac{- \bar z_m}{z_0}\Big)\,\text{Bi}'\Big(\frac{ - \bar z_m}{z_0}\Big)\>, \nonumber\\
  F'(\sigma_A - \lambda) &= \text{Bi}\Big(\frac{- \bar z_k}{z_0}\Big)\,\text{Ai}'\Big(\frac{ - \bar z_k}{z_0}\Big) - \text{Ai}\Big(\frac{- \bar z_k}{z_0}\Big)\,\text{Bi}'\Big(\frac{ - \bar z_k}{z_0}\Big)\>, 
\end{align}
as well as
\begin{align} 
  F'(\sigma_B) &= \text{Bi}\Big(\frac{- \bar z_m}{z_0}\Big)\,\text{Ai}'\Big(\frac{L - \bar z_m}{z_0}\Big) - \text{Ai}\Big(\frac{- \bar z_m}{z_0}\Big)\,\text{Bi}'\Big(\frac{L - \bar z_m}{z_0}\Big)\>, \nonumber\\
  F'(\sigma_B - \lambda) &= \text{Bi}\Big(\frac{- \bar z_k}{z_0}\Big)\,\text{Ai}'\Big(\frac{L - \bar z_k}{z_0}\Big) - \text{Ai}\Big(\frac{- \bar z_k}{z_0}\Big)\,\text{Bi}'\Big(\frac{L - \bar z_k}{z_0}\Big)\>,
\end{align}
in these Appendices.

This concludes our analytical exposition of the problem of a non-relativistic particle in a homogenous gravitational field and confined by two independently vibrating mirrors.
The subsequent numerical analysis and physical interpretation we plan to publish in a separate communication.

\section{Conclusion}\label{sec:IV}

We have analyzed the behavior of a non-relativistic particle, e.g. an ultra-cold neutron, in a homogenous gravitational field and confined between two independently harmonically vibrating mirrors. This constitutes a non-trivial generalization of the \textit{q}BOUNCE experiment.

In this paper we consider exclusively the theoretical analysis of such a generalized setup. In a subsequent publication, we plan to add our observations obtained by numerical simulations of several excitation configurations of the two vibrating mirrors. 

\section{Acknowledgments} H. A. and M. P. were both supported by the TU Wien.

\newpage 

\appendix

\section{Integrals with two Airy Functions}\label{sec:A}

\subsection{Airy Functions with the same Arguments}\label{AppA1}

We search the solution of the integral
\begin{align}
  \int_{\sigma_A}^{\sigma_B} d\sigma\,F(\sigma)\,G(\sigma)\>,
\end{align}
where $F(\sigma)$ and $G(\sigma)$ each are arbitrary linear combinations of the Airy Ai and Bi functions, i.e.
\begin{align}
  F(\sigma) &= \mathcal{C}_1\,\text{Ai}(\sigma) + \mathcal{C}_2\,\text{Bi}(\sigma)\>, \nonumber\\
  G(\sigma) &= \mathcal{C}_3\,\text{Ai}(\sigma) + \mathcal{C}_4\,\text{Bi}(\sigma)\>,
\end{align}
with arbitrary constants $\mathcal{C}_1$ to $\mathcal{C}_4$.
Following \cite{Schwinger:2013}, we differentiate
\begin{align}
  \frac{d^2}{d\sigma^2}\,G(\sigma) = \sigma\,G(\sigma)\>,
\end{align}
and find, using the prime $'$ to denote $d/d\sigma$,
\begin{align}
  \frac{d^2}{d\sigma^2}\,G'(\sigma) = G(\sigma) + \sigma\,G'(\sigma)\>.
\end{align}
Employing the two equations above we find
\begin{align}
  F(\sigma)\,G(\sigma) &= F(\sigma)\,\frac{d^2}{d\sigma^2}\,G'(\sigma) - \sigma\,F(\sigma)\,G'(\sigma) \nonumber\\
  &= F(\sigma)\,\frac{d^2}{d\sigma^2}\,G'(\sigma) - G'(\sigma)\,\frac{d^2}{d\sigma^2}\,F(\sigma) \nonumber\\
  &= \frac{d}{d\sigma}\,\Big(F(\sigma)\,\frac{d^2}{d\sigma^2}\,G(\sigma) - F'(\sigma)\,G'(\sigma)\Big) \nonumber\\  
  &= \frac{d}{d\sigma}\,\Big(\sigma\,F(\sigma)\,G(\sigma) - F'(\sigma)\,G'(\sigma)\Big)\>,
\end{align}
and finally for the integral
\begin{align}\label{AppA1}
  \int_{\sigma_A}^{\sigma_B} d\sigma\,F(\sigma)\,G(\sigma) = \sigma_B\,F(\sigma_B)\,G(\sigma_B) - \sigma_A\,F(\sigma_A)\,G(\sigma_A) - F'(\sigma_B)\,G'(\sigma_B) + F'(\sigma_A)\,G'(\sigma_A)\>.
\end{align}

\subsection{Airy Functions with the same Arguments and Operator}\label{AppA2}

Here, we search the solution of the integral
\begin{align}
  \int_{\sigma_A}^{\sigma_B} d\sigma\,F(\sigma)\,\mathcal O\,G(\sigma)\>,
\end{align}
where $\mathcal O$ is an operator, $F(\sigma)$ and $G(\sigma)$ each are arbitrary linear combinations of the Airy Ai and Bi functions again
and $\sigma_A$ and $\sigma_B$ are zeros of $F(\sigma)$, i.e. $F(\sigma_A) = F(\sigma_B) = 0$. 

We obtain the results from the corresponding results in Appendix \ref{AppA4} and by taking
the limit $\lambda\to0$. For this we need the expansions ($N = A,B$)
\begin{align}\label{AppGexp}
  G(\sigma_N - \lambda) &= G(\sigma_N) - \lambda\,G'(\sigma_N) + \frac{\lambda^2}{2}\,\sigma_N\,G(\sigma_N) - \frac{\lambda^3}{6}\,\big(G(\sigma_N) + \sigma_N\,G'(\sigma_N)\big) + \frac{\lambda^4}{24}\,\big(2G'(\sigma_N) + \sigma_N^2\,G(\sigma_N)\big) + \cdots\>, \nonumber\\  
  G'(\sigma_N - \lambda) &= G'(\sigma_N) - \lambda\,\sigma_N\,G(\sigma_N) + \frac{\lambda^2}{2}\,\big(G(\sigma_N) + \sigma_N\,G'(\sigma_N)\big) - \frac{\lambda^3}{6}\,\big(2G'(\sigma_N) + \sigma_N^2\,G(\sigma_N)\big) + \cdots\>.
\end{align}

\subsubsection{The Operator $\mathcal O = d/d\sigma$}

For the operator $\mathcal O = d/d\sigma$ we immediately find using Eq.~(\ref{AppA41})
\begin{align}
  \int_{\sigma_A}^{\sigma_B}d\sigma\,F(\sigma)\,\frac{d}{d\sigma}\,G(\sigma)  &= \lim_{\lambda\to0}\bigg\{\frac{1}{\lambda^2}\,F'(\sigma_B)\,G(\sigma_B - \lambda) - \frac{1}{\lambda^2}\,F'(\sigma_B)\,G(\sigma_B) + \frac{1}{\lambda}\,F'(\sigma_B)\,G'(\sigma_B - \lambda) \nonumber\\ 
&\quad - \frac{1}{\lambda^2}\,F'(\sigma_A)\,G(\sigma_A - \lambda) + \frac{1}{\lambda^2}\,F'(\sigma_A)\,G(\sigma_A) - \frac{1}{\lambda}\,F'(\sigma_A)\,G'(\sigma_A - \lambda)\bigg\}\>,
\end{align}
and hence using Eq.~(\ref{AppGexp})
\begin{align}\label{AppA21}
  \int_{\sigma_A}^{\sigma_B}d\sigma\,F(\sigma)\,\frac{d}{d\sigma}\,G(\sigma)  = -\frac{1}{2}\,\sigma_B\,F'(\sigma_B)\,G(\sigma_B) + \frac{1}{2}\,\sigma_A\,F'(\sigma_A)\,G(\sigma_A)\>.
\end{align}

\subsubsection{The Operator $\mathcal O = \sigma$}

For the operator $\mathcal O = \sigma$ we obtain using Eq.~(\ref{AppA42})
\begin{align}
  \int_{\sigma_A}^{\sigma_B} d\sigma\,F(\sigma)\,\sigma\,G(\sigma) &= \lim_{\lambda\to0}\bigg\{\frac{2 + \lambda^2\,\sigma_B}{\lambda^3}\,F'(\sigma_B)\,G(\sigma_B - \lambda) + \frac{2}{\lambda^2}\,F'(\sigma_B)\,G'(\sigma_B - \lambda)
 - \frac{2 + \lambda^3}{\lambda^3}\,F'(\sigma_B)\,G(\sigma_B)  \nonumber\\  
&\qquad\quad  - \frac{2 + \lambda^2\,\sigma_A}{\lambda^3}\,F'(\sigma_A)\,G(\sigma_A - \lambda)  - \frac{2}{\lambda^2}\,F'(\sigma_A)\,G'(\sigma_A - \lambda)
 + \frac{2 + \lambda^3}{\lambda^3}\,F'(\sigma_A)\,G(\sigma_A)\bigg\}\>,
 \end{align}
respectively, using Eq.~(\ref{AppGexp})
\begin{align}\label{AppA22}
  \int_{\sigma_A}^{\sigma_B} d\sigma\,F(\sigma)\,\sigma\,G(\sigma) =-\frac{1}{3}\,\sigma_B\,F'(\sigma_B)\,G'(\sigma_B) - \frac{1}{3}\,F'(\sigma_B)\,G(\sigma_B) + \frac{1}{3}\,\sigma_A\,F'(\sigma_A)\,G'(\sigma_A) + \frac{1}{3}\,F'(\sigma_A)\,G(\sigma_A)\>. 
 \end{align}

\subsubsection{The Operator $\mathcal O =  \sigma\,d/d\sigma$}

Finally, for $\mathcal O =  \sigma\,d/d\sigma$ we get using Eq.~(\ref{AppA43})
\begin{align}
  \int_{\sigma_A}^{\sigma_B} d\sigma\,F(\sigma)\,\sigma\,\frac{d}{d\sigma}\,G(\sigma) &= \lim_{\lambda\to0}\bigg\{\frac{6 + 3\lambda^2\,\sigma_B - 2\lambda^3}{\lambda^4}\,F'(\sigma_B)\,G(\sigma_B - \lambda) + \frac{6 + \lambda^2\,\sigma_B}{\lambda^3}\,F'(\sigma_B)\,G'(\sigma_B - \lambda) 
 - \frac{6}{\lambda^4}\,F'(\sigma_B)\,G(\sigma_B)  \nonumber\\  
&\qquad\quad- \frac{6 + 3\lambda^2\,\sigma_A - 2\lambda^3}{\lambda^4}\,F'(\sigma_A)\,G(\sigma_A - \lambda) - \frac{6 + \lambda^2\,\sigma_A}{\lambda^3}\,F'(\sigma_A)\,G'(\sigma_A - \lambda)  + \frac{6}{\lambda^4}\,F'(\sigma_A)\,G(\sigma_A)\bigg\}\>,
 \end{align}
and hence using Eq.~(\ref{AppGexp})
\begin{align}\label{AppA23}
  \int_{\sigma_A}^{\sigma_B} d\sigma\,F(\sigma)\,\sigma\,\frac{d}{d\sigma}\,G(\sigma) =-\frac{1}{4}\,\sigma_B^2\,F'(\sigma_B)\,G(\sigma_B) + \frac{1}{2}\,F'(\sigma_B)\,G'(\sigma_B) + \frac{1}{4}\,\sigma_A^2\,F'(\sigma_A)\,G(\sigma_A) - \frac{1}{2}\,F'(\sigma_A)\,G'(\sigma_A)\>.
 \end{align}

\subsection{Airy Functions with shifted Arguments}\label{AppA3}

Here, we give an original solution to the general integral
\begin{align}
  \int_{\sigma_A}^{\sigma_B} d\sigma\,F(\sigma)\,G(\sigma - \lambda)\>,
\end{align}
where $F(\sigma)$ and $G(\sigma)$ each are arbitrary linear combinations of the Airy Ai and Bi functions again
and $\sigma_A$ as well as $\sigma_B$ are zeros of $F(\sigma)$, i.e. $F(\sigma_A) = F(\sigma_B) = 0$.
Differentiating 
\begin{align}
  \frac{d^2}{d\sigma^2}\,F(\sigma) = \sigma\,F(\sigma)\>,
\end{align}
we find, using the prime $'$ to denote $d/d\sigma$,
\begin{align}
  \frac{d^2}{d\sigma^2}\,F'(\sigma) = F(\sigma) + \sigma\,F'(\sigma)\>.
\end{align}
Furthermore, we use 
\begin{align}
  \frac{d^2}{d\sigma^2}\,G(\sigma - \lambda) = (\sigma - \lambda)\,G(\sigma - \lambda)\>,
\end{align}
and find 
\begin{align}
  F(\sigma)\,G(\sigma - \lambda) &= G(\sigma - \lambda)\,\frac{d^2}{d\sigma^2}\,F'(\sigma) - \sigma\,G(\sigma - \lambda)\,F'(\sigma) \nonumber\\
  &= G(\sigma - \lambda)\,\frac{d^2}{d\sigma^2}\,F'(\sigma) - \frac{d^2}{d\sigma^2}\,G(\sigma - \lambda)\,F'(\sigma) - \lambda\,G(\sigma - \lambda)\,F'(\sigma) \nonumber\\
  &= \frac{d}{d\sigma}\,\Big(G(\sigma - \lambda)\,\frac{d^2}{d\sigma^2}\,F(\sigma) - G'(\sigma - \lambda)\,F'(\sigma)\Big) - \lambda\,G(\sigma - \lambda)\,F'(\sigma) \nonumber\\  
  &= \frac{d}{d\sigma}\,\Big(\sigma\,G(\sigma - \lambda)\,F(\sigma) - G'(\sigma - \lambda)\,F'(\sigma)\Big) - \lambda\,G(\sigma - \lambda)\,F'(\sigma)\>.
\end{align}
Hence
\begin{align}
  \int_{\sigma_A}^{\sigma_B} d\sigma\,F(\sigma)\,G(\sigma - \lambda) &= - G'(\sigma_B - \lambda)\,F'(\sigma_B) + G'(\sigma_A - \lambda)\,F'(\sigma_A) - \lambda\,\int_{\sigma_A}^{\sigma_B} d\sigma\,G(\sigma - \lambda)\,F'(\sigma) \nonumber\\ 
  &= - G'(\sigma_B - \lambda)\,F'(\sigma_B) + G'(\sigma_A - \lambda)\,F'(\sigma_A) - \lambda\,\frac{d}{d\lambda}\int_{\sigma_A}^{\sigma_B} d\sigma\,F(\sigma)\,G(\sigma - \lambda)\>.
\end{align}
Denoting
\begin{align}
  f(\lambda) &= \int_{\sigma_A}^{\sigma_B} d\sigma\,F(\sigma)\,G(\sigma - \lambda)\>,  \nonumber\\ 
  g(\lambda) &= - G'(\sigma_B - \lambda)\,F'(\sigma_B) + G'(\sigma_A - \lambda)\,F'(\sigma_A)\>,
\end{align}
we find the differential equation, which the integral $f(\lambda)$ has to obey, i.e.
\begin{align}\label{AppDEf}
  \lambda\,\frac{df(\lambda)}{d\lambda} + f(\lambda) = g(\lambda)\>.
\end{align}
Thereby, the solution of the integral $f(\lambda)$ can be found as solution to a differential equation.  
The homogenous solution is easily found as
\begin{align}
  f_h(\lambda) = \frac{C}{\lambda}\>,
\end{align}
with some constant $C$. For the particular solution we write
\begin{align}
  f_p(\lambda) = \frac{C(\lambda)}{\lambda}\>.
\end{align}
Using this in Eq.~(\ref{AppDEf}) we obtain
\begin{align}
  \frac{dC(\lambda)}{d\lambda} = g(\lambda)\>,
\end{align}
with the solution
\begin{align}
  C(\lambda) = C(0) + \int_0^\lambda d\lambda'\,g(\lambda')\>.
\end{align}
Finally, we find for the complete solution
\begin{align}
  f(\lambda) &= f_h(\lambda) + f_p(\lambda) \nonumber\\ 
  &= \frac{C}{\lambda} + \frac{1}{\lambda}\,\int_0^\lambda d\lambda'\,g(\lambda')\>,
\end{align}
i.e.
\begin{align}
  \int_{\sigma_A}^{\sigma_B} d\sigma\,F(\sigma)\,G(\sigma - \lambda) &= \frac{C}{\lambda} - \frac{1}{\lambda}\,F'(\sigma_B)\int_0^\lambda d\lambda'\,\frac{d}{d\sigma}\,G(\sigma - \lambda')\Big|_{\sigma=\sigma_B} + \frac{1}{\lambda}\,F'(\sigma_A)\int_0^\lambda d\lambda'\,\frac{d}{d\sigma}\,G(\sigma - \lambda')\Big|_{\sigma=\sigma_A} \nonumber\\ 
  &= \frac{C}{\lambda} + \frac{1}{\lambda}\,F'(\sigma_B)\int_0^\lambda d\lambda'\,\frac{d}{d\lambda'}\,G(\sigma_B - \lambda') - \frac{1}{\lambda}\,F'(\sigma_A)\int_0^\lambda d\lambda'\,\frac{d}{d\lambda'}\,G(\sigma_A - \lambda')\>,
\end{align}
respectively
\begin{align}
  \int_{\sigma_A}^{\sigma_B} d\sigma\,F(\sigma)\,G(\sigma - \lambda) = \frac{C}{\lambda} + \frac{1}{\lambda}\,F'(\sigma_B)\,G(\sigma_B - \lambda) - \frac{1}{\lambda}\,F'(\sigma_B)\,G(\sigma_B) - \frac{1}{\lambda}\,F'(\sigma_A)\,G(\sigma_A - \lambda) + \frac{1}{\lambda}\,F'(\sigma_A)\,G(\sigma_A)\>.  
\end{align}
The constant $C$ is obtained from the corresponding result for $\lambda=0$. We have in this limit 
\begin{align}
  \int_{\sigma_A}^{\sigma_B} d\sigma\,F(\sigma)\,G(\sigma)  &= \lim_{\lambda\to0}\frac{C}{\lambda} + F'(\sigma_B)\,\frac{d}{d\lambda}\,G(\sigma_B - \lambda)\Big|_{\lambda=0} - F'(\sigma_A)\,\frac{d}{d\lambda}\,G(\sigma_A - \lambda)\Big|_{\lambda=0} \nonumber\\  
  &= \lim_{\lambda\to0}\frac{C}{\lambda} - F'(\sigma_B)\,\frac{d}{d\sigma}\,G(\sigma)\Big|_{\sigma=\sigma_B} + F'(\sigma_A)\,\frac{d}{d\sigma}\,G(\sigma)\Big|_{\sigma=\sigma_A}\>.
\end{align}
Comparison to Eq~(\ref{AppA1}) with $F(\sigma_A) = F(\sigma_B) = 0$ provides $C=0$. Finally, we obtain
\begin{align}\label{A21}
  \int_{\sigma_A}^{\sigma_B} d\sigma\,F(\sigma)\,G(\sigma - \lambda) = \frac{1}{\lambda}\,F'(\sigma_B)\,G(\sigma_B - \lambda) - \frac{1}{\lambda}\,F'(\sigma_B)\,G(\sigma_B) - \frac{1}{\lambda}\,F'(\sigma_A)\,G(\sigma_A - \lambda) + \frac{1}{\lambda}\,F'(\sigma_A)\,G(\sigma_A)\>.  
\end{align}

\subsection{Airy Functions with shifted Arguments and Operator}\label{AppA4}

In this section, we give an original solution to the general integral
\begin{align}
  \int_{\sigma_A}^{\sigma_B} d\sigma\,F(\sigma)\,\mathcal O\,G(\sigma - \lambda)\>,
\end{align}
where $\mathcal O$ is an operator, $F(\sigma)$ and $G(\sigma)$ each are arbitrary linear combinations of the Airy Ai and Bi functions, i.e.
\begin{align}
  F(\sigma) &= \mathcal{C}_1\,\text{Ai}(\sigma) + \mathcal{C}_2\,\text{Bi}(\sigma)\>, \nonumber\\
  G(\sigma) &= \mathcal{C}_3\,\text{Ai}(\sigma) +\mathcal{C}_4\,\text{Bi}(\sigma)\>,
\end{align}
with arbitrary constants $\mathcal{C}_1$ to $\mathcal{C}_4$ and $\sigma_A$ as well as $\sigma_B$ are zeros of $F(\sigma)$, i.e. $F(\sigma_A) = F(\sigma_B) = 0$.

\subsubsection{The Operator $\mathcal{O} = d/d\sigma$}

For the operator $\mathcal{O} = d/d\sigma$ we obtain
\begin{align}
  \int_{\sigma_A}^{\sigma_B}d\sigma\,F(\sigma)\,\frac{d}{d\sigma}\,G(\sigma - \lambda)   
  &= -\frac{d}{d\lambda}\int_{\sigma_A}^{\sigma_B} d\sigma\,F(\sigma)\,G(\sigma - \lambda) \nonumber\\ 
  &=-\frac{d}{d\lambda}\left\{\frac{1}{\lambda}\,F'(\sigma_B)\,G(\sigma_B - \lambda) - \frac{1}{\lambda}\,F'(\sigma_B)\,G(\sigma_B) - \frac{1}{\lambda}\,F'(\sigma_A)\,G(\sigma_A - \lambda) + \frac{1}{\lambda}\,F'(\sigma_A)\,G(\sigma_A)\right\}\>,
\end{align}
respectively
\begin{align}\label{AppA41}
  \int_{\sigma_A}^{\sigma_B}d\sigma\,F(\sigma)\,\frac{d}{d\sigma}\,G(\sigma - \lambda)  &=
\frac{1}{\lambda^2}\,F'(\sigma_B)\,G(\sigma_B - \lambda) - \frac{1}{\lambda^2}\,F'(\sigma_B)\,G(\sigma_B) + \frac{1}{\lambda}\,F'(\sigma_B)\,G'(\sigma_B - \lambda) \nonumber\\ 
&\quad - \frac{1}{\lambda^2}\,F'(\sigma_A)\,G(\sigma_A - \lambda) + \frac{1}{\lambda^2}\,F'(\sigma_A)\,G(\sigma_A) - \frac{1}{\lambda}\,F'(\sigma_A)\,G'(\sigma_A - \lambda)\>. 
\end{align}

\subsubsection{The Operator $\mathcal{O} = \sigma$}

For the operator $\mathcal{O} = \sigma$ we find
\begin{align}
  \int_{\sigma_A}^{\sigma_B} d\sigma\,F(\sigma)\,\sigma\,G(\sigma - \lambda) 
  &= \int_{\sigma_A}^{\sigma_B} d\sigma\,F(\sigma)\left(\frac{d^2}{d\sigma^2}\,G(\sigma - \lambda) + \lambda\,G(\sigma - \lambda)\right) \nonumber\\ 
&= \left(\frac{d^2}{d\lambda^2} + \lambda\right)\int_{\sigma_A}^{\sigma_B} d\sigma\,F(\sigma)\,G(\sigma - \lambda) \nonumber\\   
&=\left(\frac{d^2}{d\lambda^2} + \lambda\right)\left\{\frac{1}{\lambda}\,F'(\sigma_B)\,G(\sigma_B - \lambda) - \frac{1}{\lambda}\,F'(\sigma_B)\,G(\sigma_B) - \frac{1}{\lambda}\,F'(\sigma_A)\,G(\sigma_A - \lambda) + \frac{1}{\lambda}\,F'(\sigma_A)\,G(\sigma_A)\right\}, 
 \end{align}
respectively
\begin{align}\label{AppA42}
  \int_{\sigma_A}^{\sigma_B} d\sigma\,F(\sigma)\,\sigma\,G(\sigma - \lambda)&=\frac{2 + \lambda^2\,\sigma_B}{\lambda^3}\,F'(\sigma_B)\,G(\sigma_B - \lambda) + \frac{2}{\lambda^2}\,F'(\sigma_B)\,G'(\sigma_B - \lambda)
 - \frac{2 + \lambda^3}{\lambda^3}\,F'(\sigma_B)\,G(\sigma_B)  \nonumber\\  
&\quad  - \frac{2 + \lambda^2\,\sigma_A}{\lambda^3}\,F'(\sigma_A)\,G(\sigma_A - \lambda)  - \frac{2}{\lambda^2}\,F'(\sigma_A)\,G'(\sigma_A - \lambda)
 + \frac{2 + \lambda^3}{\lambda^3}\,F'(\sigma_A)\,G(\sigma_A)\>. 
 \end{align}

\subsubsection{The Operator $\mathcal{O} = \sigma\,d/d\sigma$}

Finally, for $\mathcal{O} = \sigma\,d/d\sigma$ we get
\begin{align}
  &\int_{\sigma_A}^{\sigma_B} d\sigma\,F(\sigma)\,\sigma\,\frac{d}{d\sigma}\,G(\sigma - \lambda) \nonumber\\ 
  &\qquad= -\frac{d}{d\lambda}\int_{\sigma_A}^{\sigma_B} d\sigma\,F(\sigma)\,\sigma\,G(\sigma - \lambda) \nonumber\\ 
&\qquad= -\frac{d}{d\lambda}\,\bigg\{\frac{2 + \lambda^2\,\sigma_B}{\lambda^3}\,F'(\sigma_B)\,G(\sigma_B - \lambda) + \frac{2}{\lambda^2}\,F'(\sigma_B)\,G'(\sigma_B - \lambda)
 - \frac{2 + \lambda^3}{\lambda^3}\,F'(\sigma_B)\,G(\sigma_B)  \nonumber\\  
&\qquad\quad  - \frac{2 + \lambda^2\,\sigma_A}{\lambda^3}\,F'(\sigma_A)\,G(\sigma_A - \lambda)  - \frac{2}{\lambda^2}\,F'(\sigma_A)\,G'(\sigma_A - \lambda)
 + \frac{2 + \lambda^3}{\lambda^3}\,F'(\sigma_A)\,G(\sigma_A)\bigg\}\>,
\end{align}
respectively
\begin{align}\label{AppA43}
  \int_{\sigma_A}^{\sigma_B} d\sigma\,F(\sigma)\,\sigma\,\frac{d}{d\sigma}\,G(\sigma - \lambda) 
  &= \frac{6 + 3\lambda^2\,\sigma_B - 2\lambda^3}{\lambda^4}\,F'(\sigma_B)\,G(\sigma_B - \lambda) + \frac{6 + \lambda^2\,\sigma_B}{\lambda^3}\,F'(\sigma_B)\,G'(\sigma_B - \lambda) 
 - \frac{6}{\lambda^4}\,F'(\sigma_B)\,G(\sigma_B)  \nonumber\\  
&\quad- \frac{6 + 3\lambda^2\,\sigma_A - 2\lambda^3}{\lambda^4}\,F'(\sigma_A)\,G(\sigma_A - \lambda) - \frac{6 + \lambda^2\,\sigma_A}{\lambda^3}\,F'(\sigma_A)\,G'(\sigma_A - \lambda)  + \frac{6}{\lambda^4}\,F'(\sigma_A)\,G(\sigma_A)\>.
\end{align}

\section{Normalized Wavefunctions \& Matrixelements}\label{sec:B}

\subsection{Normalized Wavefunctions}

In this section, we obtain the normalized wavefunctions. They are given by
\begin{align}
  \psi^{(0)}(z) = C\,F(\sigma)\>,
\end{align}
where $F(\sigma)$ is some linear combinations of the Airy Ai and Bi functions. From the normalization 
\begin{align}
  \int_A^Bdz\,\big|\psi^{(0)}(z)\big|^2 = 1\>,
\end{align}
we obtain for the normalization constant
\begin{align}
  C = \frac{1}{\sqrt z_0}\frac{1}{\displaystyle\sqrt{\int_{\sigma_A}^{\sigma_B}d\sigma\,F^2(\sigma)}}\>,
\end{align}
and hence
\begin{align}
  \psi^{(0)}(z) = \frac{1}{\sqrt z_0}\frac{F(\sigma)}{\displaystyle\sqrt{\int_{\sigma_A}^{\sigma_B}d\sigma\,F^2(\sigma)}}\>.
\end{align}
Using Eq.~(\ref{AppA1}) we finally obtain for the normalized wavefunction
\begin{align}\label{AppNWF}
  \psi^{(0)}(z) = \frac{1}{\sqrt z_0}\frac{F(\sigma)}{\displaystyle\sqrt{F'^2(\sigma_A) - F'^2(\sigma_B)}}\>,
\end{align}
where we have used that the wavefunction vanishes at the boundaries $\sigma_A$ and $\sigma_B$, i.e. $F(\sigma_A) = F(\sigma_B) = 0$.

\subsection{Matrixelements}

Here, we obtain the matrix elements 
\begin{align}
  \int_A^B dz\,\psi_m^{(0)*}(z)\,\mathcal O\,\psi_k^{(0)}(z)\>.
\end{align}
Since $F(\sigma_A) = F(\sigma_B) = F(\sigma_A - \lambda) = F(\sigma_B - \lambda) = 0$ we obtain
\begin{align}
  \int_A^B dz\,\psi_m^{(0)*}(z)\,\mathcal O\,\psi_k^{(0)}(z) = \frac{\displaystyle\int_{\sigma_A}^{\sigma_B} d\sigma\,F(\sigma)\,\mathcal O\,F(\sigma_A - \lambda)}{\displaystyle\sqrt{F'^2(\sigma_A) - F'^2(\sigma_B)}\sqrt{F'^2(\sigma_A - \lambda) - F'^2(\sigma_B - \lambda)}}\>.
\end{align}
Explicitly, we have the relations
\begin{align}
  \sigma = \frac{z - z_m}{z_0}\>, \qquad
  \lambda = \frac{ z_k - z_m}{z_0}\>, 
\end{align}
and consequently 
\begin{align}
  \sigma_A &= \frac{A - z_m}{z_0}\>, \qquad
  \sigma_B = \frac{B -  z_m}{z_0}\>,  \nonumber\\
  \sigma_A - \lambda &= \frac{A - z_k}{z_0}\>, \qquad
  \sigma_B - \lambda = \frac{B -  z_k}{z_0}\>.
\end{align}

\subsubsection{The Operator $\mathcal O = z$}

Since $z = z_m + z_0\,\sigma$, we have for $m\neq k$ 
\begin{align}
  \int_A^B dz\,\psi_m^{(0)*}(z)\,z\,\psi_k^{(0)}(z) = z_0\,\frac{\displaystyle\int_{\sigma_A}^{\sigma_B} d\sigma\,F(\sigma)\,\sigma\,F(\sigma - \lambda)}{\displaystyle\sqrt{F'^2(\sigma_A) - F'^2(\sigma_B)}\sqrt{F'^2(\sigma_A - \lambda) - F'^2(\sigma_B - \lambda)}}\>,
\end{align}
and using Eq.~(\ref{AppA42})
\begin{align}\label{App:z1}
  \int_A^B dz\,\psi_m^{(0)*}(z)\,z\,\psi_k^{(0)}(z) = z_0\,\frac{2}{\lambda^2}\frac{\displaystyle F'(\sigma_B)\,F'(\sigma_B - \lambda) - F'(\sigma_A)\,F'(\sigma_A - \lambda)}{\displaystyle\sqrt{F'^2(\sigma_A) - F'^2(\sigma_B)}\sqrt{F'^2(\sigma_A - \lambda) - F'^2(\sigma_B - \lambda)}}\>.
\end{align}
For $m=k$ we obtain
\begin{align}
  \int_A^B dz\,\psi_m^{(0)*}(z)\,z\,\psi_m^{(0)}(z) = z_m + z_0\,\frac{\displaystyle\int_{\sigma_A}^{\sigma_B} d\sigma\,F^2(\sigma)\,\sigma}{\displaystyle F'^2(\sigma_A) - F'^2(\sigma_B)}\>,
\end{align}
and using Eqs.~(\ref{AppA1}) and (\ref{AppA22})
\begin{align}\label{App:z2}
  \int_A^B dz\,\psi_m^{(0)*}(z)\,z\,\psi_m^{(0)}(z) = z_m + z_0\,\frac{1}{3}\frac{\displaystyle\sigma_A\,F'^2(\sigma_A) - \sigma_B\,F'^2(\sigma_B)}{\displaystyle F'^2(\sigma_A) - F'^2(\sigma_B)}\>.
\end{align}

\subsubsection{The Operator $\mathcal O = d/dz$}

Since $d/dz = (1/z_0)\,d/d\sigma$, we have for $m\neq k$ 
\begin{align}
  \int_A^B dz\,\psi_m^{(0)*}(z)\,\frac{d}{dz}\,\psi_k^{(0)}(z) = \frac{1}{z_0}\frac{\displaystyle\int_{\sigma_A}^{\sigma_B} d\sigma\,F(\sigma)\,\frac{d}{d\sigma}\,F(\sigma - \lambda)}{\displaystyle\sqrt{F'^2(\sigma_A) - F'^2(\sigma_B)}\sqrt{F'^2(\sigma_A - \lambda) - F'^2(\sigma_B - \lambda)}}\>,
\end{align}
and using Eq.~(\ref{AppA41})
\begin{align}\label{App:dz1}
  \int_A^B dz\,\psi_m^{(0)*}(z)\,\frac{d}{dz}\,\psi_k^{(0)}(z) = \frac{1}{z_0}\frac{1}{\lambda}\frac{\displaystyle F'(\sigma_B)\,F'(\sigma_B - \lambda) - F'(\sigma_A)\,F'(\sigma_A - \lambda)}{\displaystyle\sqrt{F'^2(\sigma_A) - F'^2(\sigma_B)}\sqrt{F'^2(\sigma_A - \lambda) - F'^2(\sigma_B - \lambda)}}\>.
\end{align}
For $m=k$ we obtain
\begin{align}
  \int_A^B dz\,\psi_m^{(0)*}(z)\,\frac{d}{dz}\,\psi_m^{(0)}(z) &= \frac{1}{z_0}\frac{\displaystyle\int_{\sigma_A}^{\sigma_B} d\sigma\,F(\sigma)\,\frac{d}{d\sigma}\,F(\sigma)}{\displaystyle F'^2(\sigma_A) - F'^2(\sigma_B)}\>,
\end{align}
and using Eq.~(\ref{AppA21})
\begin{align}\label{App:dz2}
  \int_A^B dz\,\psi_m^{(0)*}(z)\,\frac{d}{dz}\,\psi_m^{(0)}(z) = 0\>.
\end{align}

\subsubsection{The Operator $\mathcal O =  z\,d/dz$}

Since $z\,d/dz = (z_m/z_0)\,d/d\sigma + \sigma\,d/d\sigma$, we have for $m\neq k$ 
\begin{align}
  \int_A^B dz\,\psi_m^{(0)*}(z)\,z\,\frac{d}{dz}\,\psi_k^{(0)}(z) &= \frac{z_m}{z_0}\frac{\displaystyle\int_{\sigma_A}^{\sigma_B} d\sigma\,F(\sigma)\,\frac{d}{d\sigma}\,F(\sigma - \lambda)}{\displaystyle\sqrt{F'^2(\sigma_A) - F'^2(\sigma_B)}\sqrt{F'^2(\sigma_A - \lambda) - F'^2(\sigma_B - \lambda)}} \nonumber\\
  &\quad+ \frac{\displaystyle\int_{\sigma_A}^{\sigma_B} d\sigma\,F(\sigma)\,\sigma\,\frac{d}{d\sigma}\,F(\sigma - \lambda)}{\displaystyle\sqrt{F'^2(\sigma_A) - F'^2(\sigma_B)}\sqrt{F'^2(\sigma_A - \lambda) - F'^2(\sigma_B - \lambda)}}\>,
\end{align}
and using Eqs.~(\ref{AppA41}) and (\ref{AppA43})
\begin{align}\label{App:zdz1}
  \int_A^B dz\,\psi_m^{(0)*}(z)\,z\,\frac{d}{dz}\,\psi_k^{(0)}(z) &= \frac{z_m}{z_0}\frac{1}{\lambda}\frac{\displaystyle F'(\sigma_B)\,F'(\sigma_B - \lambda) - F'(\sigma_A)\,F'(\sigma_A - \lambda)}{\displaystyle\sqrt{F'^2(\sigma_A) - F'^2(\sigma_B)}\sqrt{F'^2(\sigma_A - \lambda) - F'^2(\sigma_B - \lambda)}} \nonumber\\
  &\quad+ \frac{1}{\lambda^3}\frac{\displaystyle\big(6 + \lambda^2\,\sigma_B\big)\,F'(\sigma_B)\,F'(\sigma_B - \lambda) - \big(6 + \lambda^2\,\sigma_A\big)\,F'(\sigma_A)\,F'(\sigma_A - \lambda)}{\displaystyle\sqrt{F'^2(\sigma_A) - F'^2(\sigma_B)}\sqrt{F'^2(\sigma_A - \lambda) - F'^2(\sigma_B - \lambda)}}\>.
\end{align}
For $m=k$ we obtain
\begin{align}
  \int_A^B dz\,\psi_m^{(0)*}(z)\,z\,\frac{d}{dz}\,\psi_m^{(0)}(z) = \frac{z_m}{z_0}\frac{\displaystyle\int_{\sigma_A}^{\sigma_B} d\sigma\,F(\sigma)\,\frac{d}{d\sigma}\,F(\sigma)}{\displaystyle F'^2(\sigma_A) - F'^2(\sigma_B)} + \frac{\displaystyle\int_{\sigma_A}^{\sigma_B} d\sigma\,F(\sigma)\,\sigma\,\frac{d}{d\sigma}\,F(\sigma)}{\displaystyle F'^2(\sigma_A) - F'^2(\sigma_B)}\>,
\end{align}
and using Eqs.~(\ref{AppA21}) and (\ref{AppA23})
\begin{align}\label{App:zdz2}
  \int_A^B dz\,\psi_m^{(0)*}(z)\,z\,\frac{d}{dz}\,\psi_m^{(0)}(z) = -\frac{1}{2}\>.
\end{align}

\bibliographystyle{h-physrev3.bst}
\bibliography{DVMp}

\end{document}